\documentclass{llncs}
\usepackage{amsmath}
\usepackage{amssymb}
\usepackage{mathtools}
\usepackage{xspace}
\usepackage{tikz}
\usepackage{fixme}
\usepackage{wrapfig}
\usepackage{paralist,mdwlist}
\usepackage{manfnt,mparhack}
\usepackage[colorlinks=true, citecolor=black,linkcolor=black,urlcolor=black]{hyperref}
\usepackage{multirow}

\newcommand{\states}{\ensuremath{S} }
\newcommand{\state}{\ensuremath{s} }

\newcommand{\supp}{\ensuremath{{\sf Supp}} }

\newcommand{\weight}{\ensuremath{w} }

\newcommand{\integ}{\ensuremath{\mathbb{Z}} }
\newcommand{\strat}{\ensuremath{\sigma} }
\newcommand*{\pr}{\mathbb{P}}

\newcommand*{\expect}{\mathbb{E}}

\newcommand{\truncatedTarget}{\ensuremath{T} }
\newcommand{\truncatedSum}[1]{\ensuremath{{\sf TS}^{#1}} }

\newcommand{\nat}{\ensuremath{\mathbb{N}} }
\newcommand{\rat}{\ensuremath{\mathbb{Q}} }

\newcommand{\strats}{\ensuremath{\Sigma} }

\usetikzlibrary{shapes,arrows}

\newcommand\calD{\ensuremath{\mathcal{D}}}

\newcommand\calM{\ensuremath{\mathcal{M}}}

\newcommand{\event}{{\cal E}}

\newcommand{\sinit}{\ensuremath{s_{\sf init}}}
\usetikzlibrary{automata,calc}
\usetikzlibrary{arrows}

\title{{\bf Variations on the \\Stochastic Shortest Path Problem}\thanks{Work partially supported by ERC starting grant inVEST
	(FP7-279499) and European project CASSTING (FP7-ICT-601148).}}
\author{Mickael Randour\inst{1} \and Jean-Fran\c{}cois Raskin\inst{2} \and Ocan Sankur\inst{2}}

\institute{LSV, CNRS \& ENS Cachan, France\\
\and D\'epartement d'Informatique, Universit\'e Libre de Bruxelles (U.L.B.), Belgium}

\begin{document}
\maketitle
\begin{abstract}
In this invited contribution, we revisit the stochastic shortest path problem, and show how recent results allow one to improve over the classical solutions: 
we present algorithms to synthesize strategies with multiple guarantees on the distribution of the length of paths reaching a given target, rather than simply minimizing its expected value.
The concepts and algorithms that we propose here are applications of more general results that have been obtained recently for Markov decision processes and that are described in a series of recent papers.
\end{abstract}

\section{Introduction}

Markov decision processes (MDP)~\cite{Puterman-wiley94} are natural models for systems that exhibit both non-deterministic and stochastic evolutions. 
An MDP is executed in rounds. In each round, the MDP is in a give state and an action is chosen
by a controller (this is the resolution of non-determinism). Once this action has been fixed then the next state is determined following a probability distribution associated to the current state and the action that has been chosen by the controller.
A controller can thus be considered as a \emph{strategy} (a.k.a. \emph{policy}) that determines which action to choose according to the history of the execution so far.
MDPs have been studied intensively and there are algorithms to synthesize strategies that enforce a large variety of objectives like omega-regular 
objectives~\cite{CY-acm95}, PCTL objectives~\cite{BK-book08}, or quantitative objectives~\cite{Puterman-wiley94}.

\smallskip\noindent\textbf{One philosophy, three variants.} The classical strategy synthesis setting often considers a single objective to be optimized such as the reachability probability, or the expected cost to target. Such simple objectives are not always sufficient to describe
the properties required from an efficient controller. Indeed, on the one hand, one often has several measures of performance, and several objectives to satisfy, so the desired strategies have to settle for trade-offs between these.
On the other hand, the strategies computed in the classical setting are tailored for the precise probabilities given in the MDP, which often correspond to the \emph{average behavior} of the system in hand. 
This approach is not satisfactory if one is also interested in giving some formal guarantees under several scenarios, say, under normal conditions (i.e., average behavior), but also a minor failure, and a major failure.
In this paper, we summarize recent results that we have obtained in this direction with the common goal
of improving the strategies that can be synthesized for probabilistic systems.
They were presented in three recent publications~\cite{BFRR-stacs14,RaskinS14,RRS14}.  All three models that we studied share a common philosophy which is to provide a framework for the synthesis of strategies ensuring \textit{richer performance guarantees} than the traditional models. The three problems we tackle can be summarized as follows.

First, in~\cite{BFRR-stacs14}, we study a problem that is at the crossroad between the analysis of two-player zero-sum quantitative graph games and of quantitative MDPs. In the former, we want strategies for the controller that ensure a given minimal performance against \textit{all} possible behaviors of its environment: we ask for strict guarantees on the \textit{worst-case} performance.
In the latter, the controller plays against a stochastic environment, and we want strategies that ensure a good {expected performance}, with no guarantee on individual outcomes. Both models have clear weaknesses: strategies that are good for the worst-case may exhibit suboptimal behaviors in probable situations while strategies that are good for the expectation may be terrible in some unlikely but possible situations. The {\em beyond worst-case synthesis problem} asks to construct strategies that provide both worst-case guarantees and guarantees on the expected value against a particular stochastic model of the environment given as input. We have considered both the mean-payoff value problem and the shortest path problem. 

Second, in~\cite{RRS14}, we study multi-dimensional weighted MDPs, which are useful for modeling systems with multiple objectives. Those objectives may be conflicting, and so the analysis of trade-offs is important. To allow the analysis of those trade-offs, we study a general form of {\em percentile queries}. Percentile queries are as follows: given a multi-dimensional weighted MDP and a quantitative payoff function $f$ (such as mean-payoff or truncated sum), quantitative thresholds $v_i$ (one per dimension), and probability thresholds $\alpha_i$, we show how to compute a \textit{single} strategy that enforces that for all dimension $i$, the probability that an outcome $\rho$ satisfies $f_i(\rho) \geq v_i$ is at least $\alpha_i$.  We have obtained several new complexity results on the associated decision problems and established efficient algorithms to solve these problems.

Third, in~\cite{RaskinS14}, we introduce {\em multi-environment} MDPs (MEMDPs) which are MDPs with a \emph{set} of probabilistic transition functions. The goal in an MEMDP is to synthesize a single controller with guaranteed performances against
\emph{all} environments of this set even though the environment is unknown a priori. While MEMDPs can be seen as a special class of partially observable MDPs, several verification problems that are undecidable for partially observable MDPs, are decidable for MEMDPs and sometimes even allow for efficient solutions.

\smallskip\noindent\textbf{Stochastic shortest path.} To illustrate those results in a uniform manner, we consider the {\em stochastic shortest path problem}, {\sf SSP} problem, and study several variations. The {\em shortest path} problem is a classical optimization problem that asks, given a weighted graph, to find a path from a starting state to a target state such that the sum of weights along edges used in the path is minimized. Stochastic variants consider edges with probabilistic distributions on destinations and/or on weights. We revisit here some of those variants at the light of the results that we have obtained in the contributions described above.

\smallskip\noindent\textbf{Structure of the paper.} Our paper is organized as follows. In Sect.~\ref{sec:prelim}, we recall some elementary notions about MDPs. In Sect.~\ref{sec:sspp}, we define two classical stochastic variations on the {\sf SSP} problem: the first one asks to minimize the expected length of paths to target, and the second one asks to force short paths with high probability. In Sect~\ref{sec:bwc}, we apply the {\em beyond worst-case analysis} to the shortest path problem and summarize our results presented in~\cite{BFRR-stacs14}. In Sect.~\ref{sec:percentile}, we consider a multi-dimension version of the shortest path problem where edges both have a length and a cost. We illustrate how {\em percentile queries}, that we have studied in~\cite{RRS14}, are natural objectives for the study of trade-offs in this setting. In Sect.~\ref{sec:environment}, we study a version of the {\sf SSP} where the stochastic information is given by several probabilistic transition relations instead of one, so we apply the {\em multi-environment} MDP analysis introduced in~\cite{RaskinS14} on this variant. Throughout Sect.~\ref{sec:bwc}-\ref{sec:environment}, we also give a summary of our general results on the corresponding models, as well as additional pointers to the literature.

\section{Preliminaries}
\label{sec:prelim}

\smallskip\noindent\textbf{Markov decision processes.} A (finite) \textit{Markov decision process} (MDP) is a tuple $D=(S,s_{\sf init},A,\delta)$ where
$S$ is a finite set of \emph{states}, $s_{\sf init} \in S$ is the initial state,  $A$~is a finite set of \emph{actions}, and $\delta\colon S\times A \rightarrow \calD(S)$
is a partial function called the \emph{probabilistic transition function}, where~$\calD(S)$ denotes the set of rational probability distributions over~$S$.
The set of actions that are available in a state $\state \in \states$ is denoted by $A(s)$. 
We use $\delta(s,a,s')$ as a shorthand for $\delta(s,a)(s')$. 
A \textit{weighted} MDP $D=(S,s_{\sf init},A,\delta,\weight)$ is an MDP with a \emph{$d$-dimension integer weight function}~$w\colon A \rightarrow \mathbb{Z}^d$. For any dimension~$i \in \{1,\ldots,d\}$, we denote by $\weight_i \colon A \rightarrow \integ$ the projection of~$\weight$ to the~$i$-th dimension, $i$ is omitted when there is only one dimension.

We define a \emph{run} $\rho$ of~$D$ as a finite or infinite sequence $\rho=s_1a_1 \ldots a_{n-1}
s_n\ldots$ of states and actions such that $\delta(s_i,a_i,s_{i+1})>0$ for all~$i\geq 1$.
We denote the prefix of $\rho$ up to state $s_i$ by $\rho(i)$.
A run is called {\em initial} if it starts in the initial state $s_{\sf init}$.
We denote the set of runs of $D$ by ${\cal R}(D)$ and its set of {\em initial} runs by ${\cal R}_{s_{\sf init}}(D)$.
Finite runs that end in a state are also called \emph{histories}, and denoted by ${\cal H}(D)$ and ${\cal H}_{s_{\sf init}}(D)$, respectively.

\smallskip\noindent\textbf{Strategies.}
A~\emph{strategy}~$\sigma$ is a function ${\cal H}(D) \rightarrow \calD(A)$ such that
for all~$h \in {\cal H}(D)$ ending in~$s$, we have~$\supp(\sigma(h)) \in A(s)$, where $\supp$ denotes the support of the probability distribution. The set of all possible strategies is denoted by $\strats$.
A strategy is \emph{pure} if all histories are mapped to \emph{Dirac
distributions}.
A strategy~$\sigma$ can be encoded by a 
\emph{stochastic Moore machine}, $(\calM,\sigma_a,\sigma_u,\alpha)$ 
where~$\calM$ is a finite or infinite set of memory elements; $\sigma_a : S \times \calM \rightarrow \calD(A)$ the \emph{next action
function} where $\supp(\sigma(s,m)) \subseteq A(s)$ for any~$s\in S$ and~$m \in \calM$; $\sigma_u : A\times S \times \calM \rightarrow \calD(\calM)$ the \emph{memory update
function}; 
and~$\alpha$ the \emph{initial
distribution on~$\calM$}.
We say that~$\sigma$ is \emph{finite-memory} if~$|\calM|<\infty$, and \emph{$K$-memory} if~$|\calM|=K$;
it is memoryless if~$K=1$, thus only depends on the last state of the history. 
We define such strategies as functions $s \mapsto \calD(A(s))$ for all $s \in S$. Otherwise a strategy is \emph{infinite-memory}.

\smallskip\noindent\textbf{Markov chains.}
A weighted \textit{Markov chain} (MC) is a tuple $M=(S,d_{\sf init},\Delta,w)$ where $S$ is a (non-necessarily finite) set of \emph{states}, $d_{\sf init} \in  \calD(S)$ is the initial distribution, $\Delta \colon S \rightarrow \calD(S)$ is the \emph{probabilistic transition function}, and $w\colon S \times S \rightarrow \mathbb{Z}^d$ is a \emph{$d$-dimension weight function}. Markov chains are essentially MDPs where for all $s \in S$, we have that $\vert A(s)\vert = 1$.

We define a \emph{run} of~$M$ as a finite or infinite sequence $s_1s_2 \ldots 
s_n\ldots$ of states such that $\Delta(s_i,s_{i+1})>0$ for all~$i\geq 1$.
A run is called {\em initial} if it starts in the initial state $s$ such that $d_{\sf init}(s) > 0$.
Runs of $M$ are denoted by ${\cal R}(M)$, and its set of {\em initial} runs by ${\cal R}_{d_{\sf init}}(M)$.

\smallskip\noindent\textbf{Markov chains induced by a strategy.}
An MDP~$D=(S,s_{\sf init},A,\delta)$ and a strategy~$\sigma$ encoded by $(\calM,\sigma_a,\sigma_u,\alpha)$ determine a
Markov chain $M=D^\sigma$ defined on the state space $S \times \calM$ as follows.
The initial distribution is such that for any~$m \in \calM$,
state $(s_{\sf init},m)$ has probability $\alpha(m)$, and~$0$ for other states. For any pair of states
$(s,m)$ and~$(s',m')$, the probability of the transition $((s,m),a,(s',m'))$ is
equal to $\sigma_a(s,m)(a) \cdot \delta(s,a,s') \cdot \sigma_u(s,m,a)(m')$.
So, a~\emph{run} of~$D^\sigma$ is a finite or infinite sequence of the form
$(s_1,m_1),a_1,(s_2,m_2),a_2,\ldots$ where each
$((s_i,m_i),a_i,(s_{i+1},m_{i+1}))$ is a transition with non-zero probability
in~$D^\sigma$, and~$s_1=s_{\sf init}$. In this case, the run $s_1a_1 s_2a_2\ldots$, obtained by
projection to~$D$, is said to be \emph{compatible with~$\sigma$}.

In an MC $M$, an \textit{event} is a measurable set of runs $\event \subseteq {\cal R}_{d_{\sf init}}(M)$. Every event has a uniquely defined probability \cite{Vardi-focs85} (Carathéodory's extension theorem induces a unique probability measure on the Borel $\sigma$-algebra over ${\cal R}_{d_{\sf init}}(M)$). We denote by $\mathbb{P}_{M}(\event)$ the probability that a run belongs to $\event$ when the initial state is chosen according to $d_{{\sf init}}$, and $M$ is executed for an infinite number of steps. Given a measurable function $f \colon  {\cal R}(M) \rightarrow \mathbb{R} \cup \{\infty\}$, we denote by $\expect_{M}(f)$ the \textit{expected value} or \textit{expectation} of $f$ over initial runs in $M$.
When considering probabilities of events in $D^\sigma$, for $D$ an MDP and $\sigma$ a strategy on $D$, we often consider runs defined by their projection on~$D$. Thus, given $\event \subseteq
{\cal R}(D)$, we denote by $\pr_{D}^{\sigma}[\event]$ the probability of the runs 
of~$D^\sigma$ whose projection to~$D$ is in~$\event$.

\section{The stochastic shortest path problem}
\label{sec:sspp}

The {\em shortest path problem} in a weighted graph is a classical problem that asks, given a starting state $s$ and a set of target states $T$, to find a path from $s$ to a state $t \in T$ of minimal length (i.e., that minimizes the sum of the weights along the edges in the path). See for example~\cite{cherkassky1996shortest}. There have been several stochastic variants of this classical graph problem defined and studied in the literature, see for example~\cite{Puterman-wiley94}. We recall here two main variants of this problem, other new variants are defined and studied in the subsequent sections.

Let $D=(S,s_{\sf init},A,\delta,\weight)$ be an MDP with a \textit{single-dimensional} weight function $w\colon A \rightarrow \mathbb{N}_{0}$ that assigns to each action $a \in A$ a strictly positive integer. Let $T \subseteq S$ be a set of target states. Given an initial run $\rho=s_1 s_2 \dots s_i \dots$ in the MDP, we define its \textit{truncated sum} up to $T$ to be $\truncatedSum{T}(\rho) = \sum_{j=1}^{n-1} w(a_j)$ if $s_{n}$ is the first visit of a state in $T \subseteq \states$ within $\rho$; otherwise if $T$ is never reached, then we set $\truncatedSum{T}(\rho) = \infty$. The function $\truncatedSum{T}$ is measurable, and so this function has an expected value in a weighted MC and sets of runs defined from $\truncatedSum{T}$ are measurable.  The following two problems have been considered in the literature.

\smallskip\noindent\textbf{Minimizing the expected length of paths to target.}
Given a weighted MDP, we may be interested in strategies (choices of actions) that minimize the expected length of paths to target. This is called the {\em stochastic shortest path expectation} problem, {\sf SSP-E} for short, and it is defined as follows.

\begin{definition}[{\sf SSP-E} problem]
Given a single-dimensional weighted MDP $D=(S,s_{\sf init},A,\delta,\weight)$ and a threshold $\ell \in \mathbb{N}$, decide if there exists $\sigma$ such that $\expect^{\sigma}_{D}(\truncatedSum{T}) \leq \ell$. \end{definition}

\begin{theorem}[\cite{bertsekas1991analysis}]
\label{thm:SSPE}
The {\sf SSP-E} problem can be decided in polynomial time. Optimal pure memoryless strategies always exist and can be constructed in polynomial time.
\end{theorem}

There are several algorithms proposed in the literature to solve this problem. We recall a simple one based on linear programming (LP). For other solutions based on {\em value iteration} or {\em strategy iteration}, we refer the interested reader to, e.g.,~\cite{bertsekas1991analysis,Alfaro99}. To apply the reduction to LP, we must make the hypothesis that, for each state $s \in S$ of the MDP, there is a path from $s$ to the target set $T$. It is clear that the expectation of states that are not connected to the target set $T$ by a path is infinite. So, we will assume that all such states have first been removed from the MDP. This can easily be done in linear time. Also, it is clear that for all states in $T$, the expected length of the shortest path is trivially equal to $0$. So, we restrict our attention to states in $S \setminus T$. For each state $s \in S \setminus T$, we consider one variable $x_s$, and we define the following linear program: 

$$\max \sum_{s \in S \setminus T} x_s$$ under the constraints
$$x_s \leq \weight(a) + \sum_{s' \in S \setminus T} \delta(s,a,s') \cdot x_{s'} \mbox{~~~~~for all $s \in S \setminus T$, for all $a \in A(s)$.}$$
\noindent
It can be shown (e.g., in~\cite{bertsekas1991analysis}) that the optimal solution ${\bf v}$ for this LP is such that ${\bf v}_s$ is the expectation of the length of the shortest path from $s$ to a state in $T$ under an optimal strategy. 
 Such an optimal strategy can easily be constructed from the optimal solution ${\bf v}$. The following pure memoryless strategy $\sigma^{{\bf v}}$ is optimal: 
$$\sigma^{{\bf v}}(s)=\arg\min_{a \in A(s)} \left[ w(a) + \sum_{s' \in S \setminus T} \delta(s,a,s') \cdot {\bf v}_{s'}\right].$$

\smallskip\noindent\textbf{Forcing short paths with high probability.}
As an alternative to the expectation, given a weighted MDP, we may be interested in strategies that maximize the probability of short paths to target. 
This is called the {\em stochastic shortest path percentile} problem, {\sf SSP-P} for short,
and provides a preferable solution if we are risk-averse.
The problem is defined as follows.

\begin{definition}[{\sf SSP-P} problem]
\label{def:sspp}
Given a single-dimensional weighted MDP $D=(S,s_{\sf init},A,\delta,\weight)$, value $\ell \in \mathbb{N}$, and probability threshold $\alpha \in [0,1] \cap \mathbb{Q}$, decide if there exists a strategy $\sigma$ such that $\pr_{D}^\sigma \big[ \{ \rho \in {\cal R}_{s_{\sf init}}(D) \mid \truncatedSum{T}(\rho) \leq \ell \} \big] \geq \alpha$.
\end{definition}

\begin{theorem}
\label{thm:SSPP}
The {\sf SSP-P} problem can be decided in pseudo-polynomial time, and it is {\sf PSPACE}-hard. Optimal pure strategies with exponential memory always exist and can be constructed in exponential time. 
\end{theorem}

The {\sf PSPACE}-hardness result was recently proved in~\cite{HaaseK14}. An algorithm to solve this problem can be obtained by a (pseudo-polynomial-time) reduction to the {\em stochastic reachability problem}, {\sf SR} for short. 

Given an unweighted MDP $D=(S,s_{\sf init},A,\delta)$, a set of target states $T \subseteq S$, and a probability threshold $\alpha \in [0,1] \cap \mathbb{Q}$, the {\sf SR} problem asks to decide if there is a strategy $\sigma$ that ensures, when played from $s_{\sf init}$, to reach the set $T$ with a probability that exceeds the threshold $\alpha$. The {\sf SR} problem can also be solved in polynomial time by a reduction to linear programming. Here is a description of the LP. For all states $s \in S$, we consider a variable $x_s$ in the following LP:

$$\min \sum_{s \in S} x_s$$
\noindent
under the constraints

$$\begin{array}{lcl}
x_s=1 & \quad & \mbox{$\forall s \in T$},\\
x_s=0 & \quad &  \mbox{$\forall s \in S$ which cannot reach $T$},\\
x_s \geq \sum_{s' \in S} \delta(s,a,s') \cdot x_{s'} & \quad &  \mbox{$\forall a \in A(s)$}.
\end{array}$$
\noindent
The optimal solution ${\bf v}$ for this LP is such that ${\bf v}_s$ is the maximal probability to reach the set of targets $T$ that can be achieved from $s$. From the optimal solution ${\bf v}$, we can define a pure memoryless strategy $\sigma^{{\bf v}}$ which achieve ${\bf v}_s$ when played from $s$, we define it for all states $s \not\in T$ that can reach $T$:

$$\sigma^{{\bf v}}(s)=\arg\max_{a \in A(s)} \left[\sum_{s' \in S} \delta(s,a,s') \cdot x_{s'}\right].$$

We are now ready to define the reduction from the {\sf SSP-P} problem to the the {\sf SR} problem. Given a weighted MDP $D=(S,s_{\sf init},A,\delta,\weight)$, a set of targets $T \subseteq S$, a value $\ell \in \mathbb{N}$, and a probability threshold $\alpha \in [0,1] \cap \mathbb{Q}$, we construct an MDP $D_{\ell}$.  $D_{\ell}$ is constructed from $D$ and contains an additional information in its state space: it records the sum of the weights encountered so far. Formally, $D_{\ell}=(S',s'_{\sf init},A',\delta',w')$ where:
  \begin{itemize}
    \item $S'$ is the set of states, each one being a pair $(s,v)$, where $s \in S$ and $v \in \{ 0,1,\dots,\ell \} \cup \{ \bot \}$. Intuitively $v$ records the running sum along an execution in $D$ ($\bot > \ell$ by convention);
     \item its initial state $s'_{\sf init}$ is equal to $(s_{\sf init},0)$; 
     \item the set of actions is $A$ and the weight function is unchanged, i.e., $A'=A$ and $w'=w$;
     \item the transition relation is as follows: for all pairs $(s,v),(s',v') \in S'$, and actions $a \in A$, we have that $\delta((s,v),a)(s',v')=\delta(s,a)(s')$ if $v'=v+w(a) \leq \ell$, $\delta((s,v),a)(s',v')=\delta(s,a)(s')$ if $v'=\bot$ and $v+w(a) > \ell$, and $\delta((s,v),a)(s',v')=0$ otherwise.
  \end{itemize}
The size of $D_{\ell}$ is proportional to the size of $D$ and the value $\ell$, i.e., it is thus \textit{pseudo-polynomial} in the encoding of the {\sf SSP-P} problem.  The {\sf SR} objective in $D_{\ell}$ is to reach $T'=\{ (s,v) \mid s \in T \land v \leq \ell \}$ with a probability at least $\alpha$. 

Runs that satisfy the reachability objective in $D_{\ell}$ are in bijection with runs that reach $T$ in $D$ with a truncated sum at most $\ell$. So if there is a strategy that enforces reaching $T'$ in $D_{\ell}$ with probability $p \geq \alpha$, then there is a strategy in $D$ to ensure that $T$ is reached with a path of length at most $\ell$ with probability $p \geq \alpha$ (the strategy that is followed in $D_{\ell}$ can be followed in $D$ if we remember what is the sum of weights so far). The converse also holds. As for a reachability objective, memoryless strategies are optimal, we deduce that pseudo-polynomial-size memory is sufficient (and is sometimes necessary) in the {\sf SSP-P} problem, and the problem can be solved in pseudo-polynomial time. As the problem has been shown to be {\sf PSPACE-Hard}, this pseudo-polynomial-time solution is essentially optimal, see~\cite{HaaseK14} for details.

\begin{figure}[t]
  \centering
  \scalebox{0.88}{\begin{tikzpicture}[->,>=latex,shorten >=1pt,auto,node distance=2.5cm,bend angle=45,font=\small, inner sep=2pt]
    \tikzstyle{p1}=[draw,circle,text centered,minimum size=15mm, text width=13mm, inner sep=1pt]
    \tikzstyle{act}=[fill,circle,inner sep=1pt,minimum size=1.5pt, node distance=1cm]
    \tikzstyle{empty}=[text centered, text width=15mm]
    \node[p1] (home) at (0,0) {home};
    \node[p1] (waiting) at (-4,-4) {waiting\\room};
    \node[p1] (train) at (-2,-4) {train};
    \node[p1] (light) at (0,-4) {light\\traffic};
    \node[p1] (medium) at (2,-4) {medium\\traffic};
    \node[p1] (heavy) at (4,-4) {heavy\\traffic};
    \node[p1,double] (work) at (0,-7) {work};
    \node[act] (railway) at (-3,-2) {};
    \node[empty] at ($(railway)+(0.9,0)$) {\scriptsize railway, 2};
    \node[act] (car) at (2,-2) {};
    \node[empty] at ($(car)+(0.6,0)$) {\scriptsize car, 1};
    \node[act] (wait) at (-4,-5.9) {};
    \node[empty] at ($(wait)+(0,-0.3)$) {\scriptsize wait, 3};
    \node[act] (trainAct) at (-2,-5.5) {};
    \node[empty] at ($(trainAct)+(0.7,0)$) {\scriptsize relax, 35};
    \node[act] (back) at (-4,-1.4) {};
    \node[empty] at ($(back)+(0,0.3)$) {\scriptsize go back, 2};
    \node[act] (bike) at (6,-4) {};
    \node[empty] at ($(bike)+(0.6,0)$) {\scriptsize bike, 45};
    \node[act] (lightAct) at (0,-5.5) {};
    \node[empty] at ($(lightAct)+(0.7,0)$) {\scriptsize drive, 20};
    \node[act] (mediumAct) at (2,-5.5) {};
    \node[empty] at ($(mediumAct)+(0.7,0)$) {\scriptsize drive, 30};
    \node[act] (heavyAct) at (4,-5.5) {};
    \node[empty] at ($(heavyAct)+(0.7,0)$) {\scriptsize drive, 70};
    \path[-latex']
    (0,1.2) edge (home)
	(light) edge (lightAct)
	(lightAct) edge (work)
	(mediumAct) edge (work)
	(heavyAct) edge (work)
	(medium) edge (mediumAct)
	(heavy) edge (heavyAct)
    (waiting) edge (back)
    (back) edge (home)
    (train) edge (trainAct)
    (trainAct) edge (work)
    (home) edge (railway)
    (railway) edge  node[left,xshift=0mm]{\scriptsize $0.1$} (waiting)
    (railway) edge  node[right,xshift=0mm]{\scriptsize $0.9$} (train)
    (home) edge (car)
    (car) edge node[left,xshift=-1mm]{\scriptsize $0.2$} (light)
    (car) edge node[right,near end]{\scriptsize $0.7$} (medium)
    (car) edge node[right,xshift=1mm]{\scriptsize $0.1$} (heavy)
    (waiting) edge (wait)
    (wait) edge[bend left] node[left,xshift=0mm]{\scriptsize $0.1$} (waiting)
    ;    
	\draw [->] (home) to[out=0,in=90] (bike);
	\draw [->] (bike) to[out=270,in=0] (work);
	\draw [->] (wait) to node[right,xshift=1mm]{\scriptsize $0.9$}  (train);
  \end{tikzpicture}}
  \caption{An everyday life application of stochastic shortest path problems: choosing a mean of transport to go from home to work. Actions (black dots) are labeled with durations in minutes, and stochastic transitions are labeled with their probability.}
  \label{fig:exampleTS}
\end{figure}
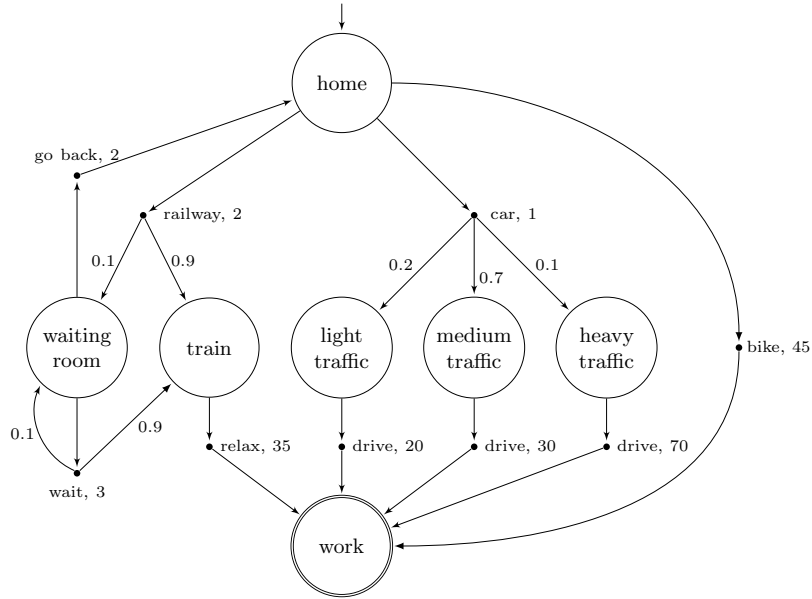

\smallskip\noindent\textbf{Illustration.}
We illustrate the concepts of this paper on a running example that we have introduced in~\cite{BFRR-stacs14} and that we extend in the subsequent sections. The MDP of Fig.~\ref{fig:exampleTS} models the choices that an employee faces when he wants to reach work from home. He has the choice between taking the train, driving or biking. When he decides to bike, he reaches his office in 45 minutes. If he decides to take his car, then the journey depends on traffic conditions that are modeled by a probabilistic distribution between light, medium and heavy traffic. The employee can also try to catch a train, which takes 35 minutes to reach work. But trains can be delayed (potentially multiple times): in that case, the employee decides if he waits or if he goes back home (and then take his car or his bike). We consider two scenarios that correspond to the two problems defined above.

If the employee wants to minimize the expected duration of his journey from home to work, we need to solve a {\sf SSP-E} problem. By Theorem~\ref{thm:SSPE}, we know that pure memoryless strategies suffice to be optimal. It turns out that in our example, taking the car is the strategy that minimizes the expected time to work: this choice gives an expectation equal to 33 minutes.  

Observe that taking the car presents some risk: if the traffic is heavy, then work is only reached after $71$ minutes. This can be unacceptable for the employee's boss if it happens too frequently.  So if the employee is {\em risk-averse}, optimizing the expectation may not be the best choice. For example, the employee may want to reach work within $40$ minutes with high probability, say $95\%$. In this case, we need to solve a {\sf SSP-P} problem. First, observe that taking the train ensures to reach work within $40$ minutes in $99\%$ of the runs. Indeed, if the train is not delayed, we reach work with $37$ minutes, and this happens with probability $9/10$. Now, if the train is late and the employee decides to wait, the train arrives in the next $3$ minutes with probability $9/10$: in that case, the employee arrives at work within $40$ minutes. So, the strategy consisting in going to the railway station and waiting for the train (as long as needed) gives us a probability $99/100$ to reach work within $40$ minutes, fulfilling our objective. Second, it is easy to see that both bicycle and car are excluded in order to satisfy the {\sf SPP-P} problem. With bicycle we reach work in $45$ minutes with probability one, and with the car we reach work in $71$ minutes with probability $1/10$, hence we miss the constraint of $40$ minutes too often.

\smallskip\noindent\textbf{Related work.} The {\sf SSP-P} problem was studied in MDPs with either all non-negative or all non-positive weights in~\cite{Ohtsubo-amc2004,SO-jcta13}. The related notion of \textit{quantile queries} was studied in~\cite{DBLP:conf/fossacs/UmmelsB13}: such queries are essentially equivalent to minimizing the value $\ell$ inside the constraint of an {\sf SSP-P} problem such that there still exists a satisfying strategy for some fixed $\alpha$. It has been recently extended to \textit{cost problems}~\cite{HaaseK14}, which can handle arbitrary Boolean combinations of inequalities over the truncated sum instead of only $\truncatedSum{T}(\rho) \leq \ell$. All those works only study single-dimensional MDPs. For the {\sf SPP-E} problem, extensions to multi-dimensional MDPs have been considered in~\cite{DBLP:conf/tacas/ForejtKNPQ11}.

\section{Good expectation under acceptable worst-case}
\label{sec:bwc}

\smallskip\noindent\textbf{Worst-case guarantees.} Assume now that the employee wants a strategy to go from home to work such that work is \textit{guaranteed} to be reached within 60 minutes (e.g., to avoid missing an important meeting with his boss). It is clear that both optimal (w.r.t.~problems {\sf SSP-E} and {\sf SSP-P} respectively) strategies of Sect.~\ref{sec:sspp} are excluded: there is the possibility of heavy traffic with the car (and a journey of $71$ minutes), and trains can be delayed indefinitely in the \textit{worst case}. 

To ensure a strict upper bound on the length of the path, an adequate model is the {\em shortest path game problem}, {\sf SP-G} for short. In a shortest path game, the uncertainty becomes adversarial: when there is some uncertainty about the outcome of an action, we do not consider a probabilistic model but we let an {\em adversary} decide the outcome of the action. So, to model a shortest path game based on an MDP $D=(S,s_{\sf init},A,\delta,\weight)$, we modify the interpretation of the transition relation as follows: after some history $h$ that ends up in state $s$, if the strategy chooses action $a \in A(s)$, then the adversary chooses the successor state within $\supp(\delta(s,a))$ without taking into account the actual values of the probabilities.  With this intuition in mind, if we fix a strategy $\sigma$ (for the controller), then the set of possible \textit{outcomes} in $D$, noted ${\sf Out}^{\sigma}_{D}$, is the set of initial runs that are compatible with $\sigma$, i.e., ${\sf Out}^{\sigma}_{D}=\left\lbrace  \rho \in {\cal R}_{s_{\sf init}}(D) \mid \forall\, i \geq 0 \colon\: a_i \in \supp(\sigma(\rho(i))) \right\rbrace$.
\noindent Now, we can define the {\sf SP-G} problem as follows.

\begin{definition}[{\sf SP-G} problem]
Given single-dimensional weighted MDP $D = (S,s_{\sf init},A,\delta,\weight)$, set of target states $T \subseteq S$, and value $\ell \in \mathbb{N}$, decide if there exists a strategy $\sigma$ such that for all $\rho \in {\sf Out}^{\sigma}_{D}$, we have that $\truncatedSum{T}(\rho) \leq \ell$.
\end{definition}

\begin{theorem}[\cite{DBLP:dblp_journals/mst/KhachiyanBBEGRZ08}]
The {\sf SP-G} problem can be decided in polynomial time. Optimal pure memoryless strategies always exist and can be constructed in polynomial time. 
\end{theorem}

Under the hypothesis that actions in $D$ have strictly positive weight, 
the controller has no interest in forming cycles, and if he cannot avoid to close cycles (before reaching $T$), then there will be outcomes that will never reach $T$, yielding an infinite truncated sum.  As a consequence, the only option for the controller is to win within $|S|=n$ steps.  So, to solve the {\sf SP-G} problem, we compute for each state $s$ and for each $i$, $0 \leq i \leq n$, the value $\mathbb{C}(s,i)$, representing the lowest bound on the length to the target $T$ from $s$ that the controller can ensure, if the game is played for $i$ steps. Those values can be computed using {\em dynamic programming} as follows: for all $s \in T$, $\mathbb{C}(s,0)=0$, and for all $s \in S \setminus T$, $\mathbb{C}(s,0)=+\infty$. Now, assume that $0 < i < n$ and that we have already computed $\mathbb{C}(s,i-1)$ for all $s \in S$. Then for $i$ steps, we have that 
$$\mathbb{C}(s,i) = \min\Big[\mathbb{C}(s,i-1),\: \min_{a \in A(s)} \max_{s' \in \supp(\delta(s,a))} w(a)+\mathbb{C}(s',i-1)\Big].$$
So, $\mathbb{C}(s_{\sf init},n)$ can be computed in polynomial time, and we have that the controller can force to reach $T$ from $s_{\sf init}$ with a path of length at most $\ell$ if and only if  $\mathbb{C}(s_{\sf init},n) \leq \ell$.

\smallskip\noindent\textbf{Related work.} For results about the {\sf SP-G} problem when weights can also be negative, we refer the interested reader to~\cite{BrihayeGHM14} where a pseudo-polynomial-time algorithm has been designed and to~\cite{FiliotGR12} where complexity issues are discussed (see Theorem~8 in that reference). In multi-dimensional MDPs with both positive and negative weights, it follows from results on total-payoff games that the {\sf SP-G} problem is undecidable~\cite{DBLP:conf/atva/Chatterjee0RR13}.

\smallskip\noindent\textbf{Illustration.}
If we apply this technique on the example of Fig.~\ref{fig:exampleTS}, it shows that taking bicycle is a safe option to ensure the strict $60$ minutes upper bound. However, the expected time to reach work when following this strategy is $45$ minutes, which is far from the optimum of $33$ minutes that can be obtained when we neglect the worst-case constraint.

In answer to this, we may be interested in synthesizing a strategy that minimizes the expected time to work under the constraint that work is reached within $60$ minutes in the worst case. We claim that the optimal strategy in this case is the following: try to take the train, if the train is delayed three times consecutively, then go back home and take the bicycle. This strategy is safe as it always reaches work within 58 minutes and its expectation is $\approx 37,34$ minutes (so better than taking directly the bicycle).
Observe that it is pure but requires finite memory, in contrast to the case of problems {\sf SSP-E} and {\sf SSP-G}.

\smallskip\noindent\textbf{Beyond worst-case synthesis.} In~\cite{BFRR-stacs14,DBLP:journals/corr/BruyereFRR14}, we study the synthesis of strategies that ensure, \textit{simultaneously}, a worst-case threshold (when probabilities are replaced by adversarial choices), and a good expectation (when probabilities are taken into account).
We can now recall the precise definition of the problem.

\begin{definition}[{\sf SSP-WE} problem]
Given a single-dimensional weighted MDP $D=(S,s_{\sf init},A,\delta,\weight)$, a set of target states $T \subseteq S$, and two values $\ell_1,\ell_2 \in \mathbb{N}$, decide if there exists a strategy $\sigma$ such that:
  \begin{enumerate}
  	\item $\forall\,\rho \in {\sf Out}^{\sigma}_{D}\colon\:  \truncatedSum{T}(\rho) \leq \ell_1$,
	\item $\expect^{\sigma}_{D}(\truncatedSum{T}) \leq \ell_2$.
  \end{enumerate}
\end{definition}

While the {\sf SP-G} problem and the {\sf SSP-E} problem are both solvable in polynomial time and pure memoryless strategies suffice in both cases, the {\sf SSP-WE} problem proves to be inherently harder.

\begin{theorem}[\cite{BFRR-stacs14}]
The {\sf SSP-WE} problem can be decided in pseudo-polynomial time and is {\sf NP}-hard. Pseudo-polynomial memory is always sufficient and in general necessary, and satisfying strategies can be constructed in pseudo-polynomial time. 
\end{theorem}

The algorithm proposed in~\cite{BFRR-stacs14} to solve the {\sf SSP-WE} problem can be summarized as follows. First, construct the MDP $D_{\ell}$ as for solving the {\sf SSP-P} problem. States of $D_{\ell}$ are pairs $(s,v)$ where $s \in S$ is a state of $D$ and $v$ is the sum of weights of edges traversed so far. Consider the target $T' = \{ (s,v) \mid s \in T \land v \leq \ell \}$. Second, compute for each state $(s,v)$ what are the {\em safe actions}, noted $\mathbb{A}(s,v)$, that ensure to reach $T'$ in $D_{\ell}$ no matter how the adversary resolves non-determinism.  
$\mathbb{A}(s,v)$ can be computed inductively as follows: we start with $\mathbb{A}_0(s,v)=A(s)$ if $v \leq \ell$ and $\mathbb{A}_0(s,v)=\emptyset$ if $v=\bot$, i.e., a priori, all the actions are good in states that have not yet exceeded the sum $\ell$ while states that have exceeded $\ell$ are hopeless and none of the actions are good. Assume that we have computed $\mathbb{A}_i(s,v)$, for $i \geq 0$, then $\mathbb{A}_{i+1}(s,v)=\{ a \in \mathbb{A}_{i}(s,v) \mid \forall\, (s',v') \in \supp(\delta((s,v),a))\colon\: \mathbb{A}_{i}(s,v) \not=\emptyset\}$. As the set of good actions is finite and is decreasing, it is easy to see that this process ends after a finite number of steps that is polynomial in the size of $D_{\ell}$. We note $D_{\ell}^{\mathbb{A}}$, the MDP $D_{\ell}$ limited to the safe actions. Then, it remains to solve the {\sf SSP-E} on $D_{\ell}^{\mathbb{A}}$. The overall complexity of the algorithm is pseudo-polynomial, and the {\sf NP}-hardness result established in~\cite{BFRR-stacs14} implies that we cannot hope to obtain a truly-polynomial-time algorithm unless ${\sf P}={\sf NP}$.

\smallskip\noindent\textbf{Additional results.}
In~\cite{BFRR-stacs14,DBLP:journals/corr/BruyereFRR14}, we also study the so-called beyond worst-case synthesis for models with the \textit{mean-payoff} function instead of the truncated sum. Mean-payoff games~\cite{em79} are infinite-duration, two-player
zero-sum games played on weighted graphs. In those games, the controller wants to maximize the long-run average of the weights of the edges traversed during the game while the adversary aims to minimize this long-run average. Given a mean-payoff game and a stochastic model of the adversary, their product defines an MDP on which we study the problem {\sf MP-WE}, the mean-payoff analogue of problem {\sf SSP-WE}.
We have shown that it is in ${\sf NP} \cap {\sf coNP}$ for {\em finite-memory} strategies, essentially matching the complexity of the simpler problem {\sf MP-G} of solving mean-payoff games without considering the expected value. We have also established that pure strategies with pseudo-polynomial-memory are sufficient. Our synthesis algorithm is much more complex than for {\sf SSP-WE}, and requires to overcome several technical difficulties to prove ${\sf NP} \cap {\sf coNP}$-membership. 

\section{Percentile queries in multi-dimensional MDPs}
\label{sec:percentile}

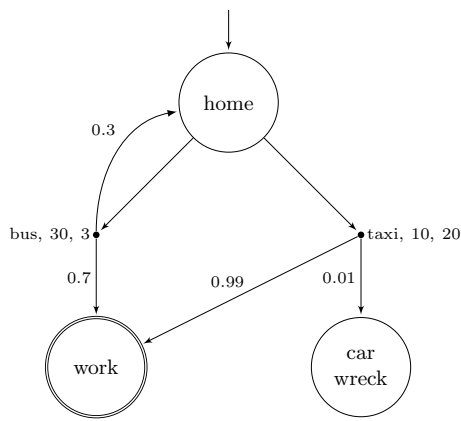
\begin{wrapfigure}{r}{65mm}
  \centering
  \scalebox{0.88}{\begin{tikzpicture}[->,>=latex,shorten >=1pt,auto,node distance=2.5cm,bend angle=45,font=\small, inner sep=2pt]
    \tikzstyle{p1}=[draw,circle,text centered,minimum size=15mm, text width=13mm, inner sep=1pt]
    \tikzstyle{act}=[fill,circle,inner sep=1pt,minimum size=1.5pt, node distance=1cm]
    \tikzstyle{empty}=[text centered, text width=15mm]
    \node[p1] (home) at (0,0) {home};
    \node[p1,double] (work) at (-2,-4) {work};
    \node[p1] (wreck) at (2,-4) {car\\wreck};
    \node[act] (bus) at (-2,-2) {};
    \node[empty] at ($(bus)+(-0.7,0)$) {\scriptsize bus, 30, 3};
    \node[act] (taxi) at (2,-2) {};
    \node[empty] at ($(taxi)+(0.8,0)$) {\scriptsize taxi, 10, 20};
    \path[-latex']
    (0,1.4) edge (home)
    (home) edge (bus)
    (bus) edge node[left,xshift=0mm,yshift=0mm]{\scriptsize $0.7$} (work)
    (home) edge (taxi)
    (taxi) edge node[left,xshift=0mm,yshift=1.4mm]{\scriptsize $0.99$} (work)
    (taxi) edge node[left,xshift=0mm,yshift=0mm]{\scriptsize $0.01$} (wreck)
    ;    
	\draw [->] (bus) to[out=90,in=190] node[above,xshift=-2mm,yshift=2mm]{\scriptsize $0.3$} (home);
  \end{tikzpicture}}
  \caption{Multi-percentile queries can help when actions both impact the duration of the journey (first dimension) and its cost (second dimension): trade-offs have to be considered.}
  \label{fig:percentile}
  \vspace{-5mm}
\end{wrapfigure}

\smallskip\noindent\textbf{Illustration.} Consider the MDP $D$ depicted in Fig.~\ref{fig:percentile}. It gives a simplified choice model for commuting from home to work, but introduces two-dimensional weights: each action is labeled with a duration, in minutes, and a cost, in dollars. Multi-dimensional MDPs are useful to analyze systems with {\em multiple objectives} that are potentially conflicting and make necessary the analysis of trade-offs. For instance, we may want a choice of transportation that gives us high probability to reach work in due time but also limits the risk of an expensive journey. Since faster options are often more expensive, trade-offs have to be considered.

Recall the {\sf SSP-P} problem presented in Def.~\ref{def:sspp}: it asks to decide the existence of strategies satisfying a \textit{single percentile constraint}. This problem can only be applied to single-dimensional MDPs. For example, one may look for a strategy that ensures that $80\%$ of compatible initial runs take at most $40$ minutes (constraint C1), \textit{or} that $50\%$ of them cost at most $10$ dollars (C2). A good strategy for C1 would be to take the taxi, which guarantees that work is reached within $10$ minutes with probability $0.99 > 0.8$. For C2, taking the bus is a good option, because already $70\%$ of the runs will reach work for only $3$ dollars. Note that taking the taxi does not satisfy C2, nor does taking the bus satisfy C1.

In practice, a desirable strategy should be able to satisfy \textit{both} C1 and C2. This is the goal of our model of \textit{multi-constraint percentile queries}, introduced in~\cite{RRS14}. For example, an appropriate strategy for the conjunction (C1 $\wedge$ C2) is to try the bus once, and then take the taxi if the bus does not depart. Indeed, this strategy ensures that work is reached within $40$ minutes with probability larger than $0.99$ thanks to runs home$\cdot$bus$\cdot$work (probability $0.7$ and duration $30$) and home$\cdot$bus$\cdot$home$\cdot$taxi$\cdot$work (probability $0.297$ and duration $40$). Furthermore, it also ensures that more than half the time, the total cost to target is at most $10$ dollars, thanks to run home$\cdot$bus$\cdot$work which has probability $0.7$ and cost $3$. Observe that this strategy requires \textit{memory}. In this particular example, it is possible to build another acceptable strategy which is memoryless but requires \textit{randomness}. Consider the strategy that flips an unfair coin in home to decide if we take the bus or the taxi, with probabilities $3/5$ and $2/5$ respectively. Constraint C1 is ensured thanks to runs home$\cdot$bus$\cdot$work (probability $0.42$) and home$\cdot$taxi$\cdot$work (probability $0.396$). Constraint C2 is ensured thanks to runs (home$\cdot$bus)$^{n}\cdot$work with $n = 1, 2, 3$: they have probabilities $0.42$, $\geq 0.07$ and $\geq 0.01$ respectively, totaling to $\geq 0.5$, while they all have cost at most $3\cdot 3 = 9 < 10$. As we will see, percentile queries in general require strategies that both use memory \textit{and} randomness, in constrast to the previous problems which could forgo randomness.

\smallskip\noindent\textbf{Percentile queries.} In~\cite{RRS14}, we study the synthesis of strategies that enforce percentile queries for the shortest path.

\begin{definition}[{\sf SSP-PQ} problem]
Given a $d$-dimensional weighted MDP $D=(S,s_{\sf init},A,\delta,\weight)$, and $q \in \nat$ percentile constraints described by sets of target states $T_{i} \subseteq S$, dimensions $k_{i} \in \{1, \ldots{}, d\}$, value thresholds $\ell_{i} \in \nat$ and probability thresholds $\alpha_{i} \in [0,1] \cap \rat$, where $i \in \{1, \ldots{}, q\}$, decide if there exists a strategy $\sigma$ such that
\begin{equation*}
\forall\, i\in \{1, \ldots{}, q\},\; \pr_{D}^\strat\big[\truncatedSum{\truncatedTarget_{i}}_{k_{i}} \leq \ell_i\big] \geq
\alpha_i,
\end{equation*}
where $\truncatedSum{\truncatedTarget_{i}}_{k_{i}}$ denotes the truncated sum on dimension $k_{i}$ and w.r.t.~target set $\truncatedTarget_{i}$.
\end{definition}
Our algorithm is able to solve the problem for queries with multiple constraints, potentially related to different dimensions of the weight function and to different target sets: this offers great flexibility which is useful in modeling applications.

\begin{theorem}[\cite{RRS14}]
The {\sf SSP-PQ} problem can be decided in exponential time in general, and pseudo-polynomial time for single-dimension single-target multi-contraint queries. The problem is {\sf PSPACE}-hard even for single-constraint que\-ries. Randomized exponential-memory strategies are always sufficient and in general necessary, and satisfying strategies can be constructed in exponential time.
\end{theorem}

The first step to solve an {\sf SSP-PQ} problem on MDP $D$ is to build a new MDP $D_{\ell}$ similarly to what was defined for the {\sf SSP-P} problem, but with $\ell = \max_{i} \ell_{i}$, and adapting the construction to multi-dimensional weights. In particular, we observe that a run can only be disregarded when the sum on each of its dimensions exceeds $\ell$. Essentially, some runs may satisfy only a subset of constraints and still be interesting for the controller, as seen in the example above.
Still, the size of $D_{\ell}$ can be maintained to a \textit{single}-exponential by defining a suitable equivalence relation between states (pseudo-polynomial for single-dimensional MDPs and single-target queries).
Precisely, 
the states of~$D_{\ell}$ are in $S \times \left(\{0,\ldots,\ell\} \cup \{\bot\}\right)^d$. 
Now, for each constraint $i$, we compute a set of target states $R_{i}$ in $D_{\ell}$ that exactly captures all runs satisfying the inequality of the constraint. 

We are left with a \textit{multiple reachability problem} on~$D_{\ell}$: we look for a strategy~$\strat_{\ell}$ that ensures that each of these sets $R_{i}$ is reached with probability $\alpha_{i}$. This is a generalization of the {\sf SR} problem defined above. It follows from~\cite{EKVY-lmcs08} that this multiple reachability problem can be answered in time polynomial in $\vert D_{\ell}\vert$ but exponential in the number of sets $R_{i}$, i.e., in $q$. The complexity can be reduced for single-dimensional MDPs and queries with a unique target $T$: in that case, sets $R_{i}$ can be made absorbing, and the multiple reachability problem can be answered in time polynomial in $D_{\ell}$ through linear programming. Overall, our algorithm thus requires pseudo-polynomial time in that case. It is clear that $\strat_{\ell}$ can be easily translated to a good strategy $\strat$ in $D$ and conversely.

The {\sf PSPACE}-hardness result already holds for the single-constraint case, i.e., the {\sf SSP-P} problem (Theorem~\ref{thm:SSPP}), following results of~\cite{HaaseK14}. Hence the {\sf SSP-PQ} framework offers a wide extension for basically no price in decision complexity.

\smallskip\noindent\textbf{Additional results.} In~\cite{RRS14}, we establish that the {\sf SSP-PQ} problem becomes undecidable if we allow for both negative and positive weights in multi-dimensional MDPs, even with a unique target set.

Furthermore, in~\cite{RRS14}, we study the concept of percentile queries for a large range of classical payoff functions, not limited to the truncated sum: \textit{sup}, \textit{inf}, \textit{limsup}, \textit{liminf}, \textit{mean-payoff} and \textit{discounted sum}. In all cases, the complexity for the most general setting - multi-dimensional MDPs, multiple constraints - is at most exponential, better in some cases. Interestingly, when the query size is fixed, all problems except for the discounted sum can be solved in polynomial time. Note that in most applications, the query size can be reasonably bounded while the model can be very large, so this framework is ideally suited. In many cases, we show how to reduce the complexity for single-dimensional queries, and for single-constraint queries.
We also improve the knowledge of the multiple reachability problem sketched above by proving its {\sf PSPACE}-hardness and identifying the subclass of queries with nested targets as solvable in polynomial time.

\smallskip\noindent\textbf{Related work.} As mentioned in Sect.~\ref{sec:sspp}, there are several works that extend the {\sf SSP-P} problem in different directions. In particular, \textit{cost problems}, recently introduced in~\cite{HaaseK14}, can handle arbitrary Boolean combinations of inequalities~$\varphi$ over the truncated sum inside an {\sf SSP-P} problem: it can be written as $\pr_{D}^\strat\big[\truncatedSum{\truncatedTarget}  \models \varphi\big] \geq
	\alpha$. Observe that this is orthogonal to our percentile queries. Cost problems are studied on single-dimensional MDPs and all the inequalities relate to the same target $\truncatedTarget$, in contrast to our setting which allows both for multiple dimensions and multiple target sets. The single probability threshold bounds the probability of the whole event~$\varphi$ whereas we analyze each event independently. Both settings are in general incomparable (see~\cite{RRS14}), but they share the {\sf SSP-P} problem as a common subclass.

\section{Multiple environments}
\label{sec:environment}
The probabilities in a stochastic process represent a model of the \emph{environment}. For instance,
in Fig.~\ref{fig:exampleTS}, the probability of a train coming when we wait in the train station is a simplified model of the behavior of the train network.
Clearly, this behavior can be significantly different on some particular days, for instance, when there is a strike.
In this section, we consider the problem of synthesizing strategies in probabilistic systems with guarantees against a finite number of different \emph{environments}.

\smallskip\noindent\textbf{Illustration.}
Let us consider again the problem of commuting to work, and assume that some days there may be an unannounced strike (S) in the train service,
and an accident (A) in the highway. Thus, four settings are possible: (), (A), (S), (AS).
When there is a strike, there is no train service; and when there is an accident, the highway is blocked. We assume that
we are not informed of the strike or the accident in advance.
Our goal is to synthesize a strategy with guarantees against these four environments with \emph{no prior knowledge} of the situation we are in.

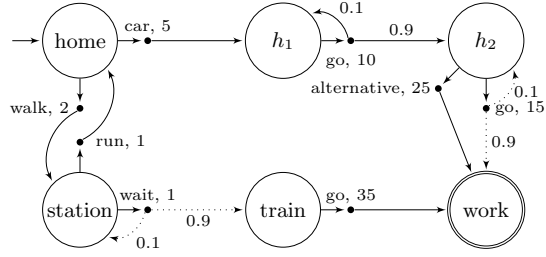
\begin{figure}[t]
  \centering
  \scalebox{0.9}{\begin{tikzpicture}[->,>=latex,shorten >=1pt,auto,node distance=2.5cm,bend angle=45,font=\small, inner sep=2pt]
    \tikzstyle{p1}=[draw,circle,text centered,minimum size=2mm, text width=10mm, inner sep=1pt]
    \tikzstyle{act}=[fill,circle,inner sep=1pt,minimum size=1.5pt, node distance=1cm]
    \tikzstyle{empty}=[]
    \node[p1] (home) at (0,0) {home};
    \node[act,below  of=home] (hwalk) {};
    \node at ($(hwalk)+(-0.6,0)$) {\scriptsize walk, 2};
    \node[act,right  of=home] (car) {};
    \node at ($(car)+(0,0.2)$) {\scriptsize car, 5};
    \node[p1,below of=home] (st) {station};
    \node[act,above  of=st] (swalk) {};
    \node at ($(swalk)+(0.6,0)$) {\scriptsize run, 1};
    \node[act,right of=st] (swait) {};
    \node at ($(swait)+(0,0.2)$) {\scriptsize wait, 1};
    \node[p1,right of=st, node distance=3cm] (tr) {train};
    \node[act,right of=tr] (departs) {};
    \node at ($(departs)+(0,0.2)$) {\scriptsize go, 35};
    \node[p1,double,right of=tr, node distance=3cm] (work) {work};
    \node[p1,right of=home, node distance=3cm] (h1) {$h_1$};
    \node[act,right of=h1] (h1go){};
    \node at ($(h1go)+(0,-0.3)$) {\scriptsize go, 10};
    \node[p1,right of=h1, node distance=3cm] (h2) {$h_2$};
    \node[act,below of=h2] (h2go){};
    \node at ($(h2go)+(0.5,0)$) {\scriptsize go, 15};
    \node[act, below left of=h2] (h2dev){};
    \node at ($(h2dev)+(-1,0)$) {\scriptsize alternative, 25};
    \path[-latex']
    (-1,0) edge (home)
    (home) edge (hwalk)
    (hwalk) edge[bend right] (st)
    (swalk) edge[bend right] (home)
    (st) edge (swalk)
    (st) edge (swait)
    (swait) edge[dotted, bend left] node[below right]{\scriptsize $0.1$} (st)
    (swait) edge[dotted] node[below]{\scriptsize $0.9$} (tr)
    (tr) edge (departs)
    (departs) edge (work)
    (home) edge (car)
    (car) edge (h1)
    (h1) edge (h1go)
    (h1go) edge[bend right] node[above right]{\scriptsize $0.1$} (h1)
    (h1go) edge node[above]{\scriptsize $0.9$} (h2)
    (h2) edge (h2go)
    (h2go) edge[dotted, bend right] node[right]{\scriptsize $0.1$} (h2)
    (h2go) edge[dotted]  node[right]{\scriptsize $0.9$}(work)
    (h2) edge (h2dev)
    (h2dev) edge (work);    
  \end{tikzpicture}}
  \caption{Commuting to work. States $h_1,h_2$ represent sections of the highway.
  After~$h_1$, one may take an alternative road which is longer but  not affected by traffic.
}
  \label{fig:memdp-commute}
\vspace{-4mm}
\end{figure}

Consider the MDP~$D$ of Fig.~\ref{fig:memdp-commute}, which models the \emph{normal conditions} without strike or accident.
We will define three different MDPs from~$D$ on the same state space to model the three other environments, by modifying the probabilities of dotted edges.
For each environment $E \in \{(A),(S),(AS)\}$, we define MDP $D^E$ from $D$ as follows.
\begin{enumerate}
\item For $D^{(S)}$, action {\scriptsize wait} from state {\scriptsize stat} deterministically leads back to {\scriptsize stat}.
\item For $D^{(A)}$, action {\scriptsize go} from state {\scriptsize $h_2$} deterministically leads back to {\scriptsize $h_2$}.
\item For $D^{(AS)}$, we apply both items 1 and 2.
\end{enumerate}
Note that if strikes and accidents have small probabilities, instead of creating separate models, one could integrate their effect
in a single model by adjusting the probabilities in~$D$, for instance, by reducing the probability of moving forward in the highway.
Such an approach may be useful (and simpler) for an \emph{average analysis}. However, we are interested here in giving guarantees against each scenario
rather than optimizing a global average. 
Our formulation can rather be modeled by \emph{partially observable} MDPs since 
the strategy is not aware of the state of the system. However, most problems
are undecidable in this setting~\cite{ChatterjeeCT13}.

Our objective is to get to work with high probability within reasonable time. More precisely,
we would like to make sure to be at work, with probability $0.95$ in all cases: in 40 minutes if there is no strike,
in 50 minutes if there is a strike but no accident, and 75 minutes if there is a strike and an accident.
More formally, we would like to synthesize a \emph{single} strategy~$\sigma$ such that:

\vspace{1mm}
\begin{minipage}[b]{0.45\linewidth}
\centering
  \begin{itemize}
	\item $\pr_{D}^\sigma[\truncatedSum{T} \leq 40]\geq 0.95$, 
	\item $\pr_{D^{(S)}}^\sigma[\truncatedSum{T} \leq 50]\geq 0.95$, 
  \end{itemize}
\end{minipage}
\hspace{0.5cm}
\begin{minipage}[b]{0.45\linewidth}
\centering
  \begin{itemize}
	\item $\pr_{D^{(A)}}^\sigma[\truncatedSum{T} \leq 40]\geq 0.95$, 
	\item $\pr_{D^{(SA)}}^\sigma[\truncatedSum{T} \leq 75]\geq 0.95$.
  \end{itemize}
\end{minipage}

\smallskip\noindent\textbf{Solution.}
We will describe a strategy that satisfies our objective.
First, note that we shouldn't take the car right away since even if we take the alternative road,
we will be at work in $40$ minutes only with probability $0.90$ (even if there is no accident, 
we may spend $20$ minutes in~$h_1$).
Our strategy is the following. We first walk to the train station, and wait there at most twice. Clearly, if there is no strike,
we get to work in less than 40 minutes with probability at least $0.99$. Otherwise, we run back home, and take the car.
Note that we already spent 5 minutes at this point. Our strategy on the highway is the following. 
We take the alternative road if, and only if we failed to make progress twice by taking action go
(e.g., we observed $h_1\cdot\text{go}\cdot h_1\cdot\text{go}\cdot h_2\cdot\text{go}\cdot h_2$).

We already saw that in the absence of strike, this strategy satisfies our objective.
If there is a strike but no accident, we will surely take the car.
Then the history ending with $h_1\cdot \text{go}\cdot h_2\cdot \text{go}\cdot\text{work}$ has probability $0.81$ and takes $30$ minutes.
The histories ending with $h_1\cdot \text{go}\cdot h_1 \cdot \text{go} \cdot h_2\cdot \text{go}\cdot\text{work}$ 
and $h_1 \cdot \text{go} \cdot h_2 \cdot \text{go}\cdot h_2\cdot \text{go}\cdot \text{work}$
have each probability $0.081$ and take $40$ and $45$ minutes respectively. Overall, with probability at least $0.97$ we
get to work in at most $50$ minutes.
If there is a strike and an accident, then the history $h_2 \cdot \text{go} \cdot \text{work}$ is never observed.
In this case, the history ending with $h_1 \cdot \text{go} \cdot h_2 \cdot \text{go} \cdot h_2 \cdot \text{go} \cdot h_2\cdot \text{alternative}$ has
probability $0.90$ and takes $75$ minutes, and history $h_1 \cdot \text{go} \cdot h_1 \cdot \text{go} \cdot h_2 \cdot \text{go} \cdot h_2 \cdot \text{alternative}$ has
probability $0.09$ and takes $75$ minutes. Hence we ensure the constraint with probability $0.99$.

\smallskip\noindent\textbf{Algorithms.}
Formally, we define a \emph{multi-environment MDP} as a tuple 
$D=\big(S,\sinit,A,(\delta_i)_{1\leq i\leq k}, (w_i)_{1\leq i\leq k}\big)$,
where each $(S,\sinit,A,\delta_i,w_i)$ is an MDP, corresponding to a different environment.

\begin{definition}[{\sf SSP-ME} problem]
  Given any single-dimensional multi-envi\-ron\-ment MDP~$D=(S,\sinit,A,(\delta_i)_{1\leq i\leq k},(w_i)_{1\leq i\leq k}\big)$,
  target states $T \subseteq S$, thresholds $\ell_1,\ldots,\ell_k \in \mathbb{N}$,
  and probabilities~$\alpha_1,\ldots,\alpha_k \in [0,1]\cap \rat$, 
  decide if there exists a strategy~$\sigma$
  satisfying
  \[\forall i \in \{1,\ldots,k\},\: \pr_{D_{i}}^\sigma[ \truncatedSum{T}\leq \ell_i ] \geq \alpha_i.\]
\end{definition}

For the particular case of $\alpha_1=\ldots=\alpha_k=1$, the problem is called
the \emph{almost-sure} {\sf SSP-ME} problem.
The \emph{limit-sure} {\sf SSP-ME} problem asks whether the {\sf SSP-ME} problem has a solution
for \emph{all} probability vectors $(\alpha_1,\ldots,\alpha_k) \in ]0,1[^k$.
If the limit-sure problem can be satisfied, the almost-sure case can be approximated arbitrarily closely.
Note that in some multi-environment MDPs, the limit-sure {\sf SSP-ME} problem has a solution
although the almost-sure one does not.

\begin{theorem}[\cite{RaskinS14}]
  The almost-sure and limit-sure {\sf SSP-ME} problems can be solved in pseudo-polynomial time
  for a fixed number of environments.
  Finite memory suffices for the al\-most-sure case, and a family of finite-memory strategies
  that witnesses the limit-sure problem can be computed.  
\end{theorem}

We analyze the structure
of the MDPs to identify \emph{learning components} in which one can almost-surely (resp. limit-surely) 
determine the current environment. Once these are identified, one can
transform the MDPs into simpler forms on which known algorithms on (single-environment) MDPs are applied~\cite{RS-fsttcs14}.

For an example of a \emph{learning component}, consider two states $s,t$ and action~$a$,
with $\delta_1(s,a,t) = 0.9, \delta_1(s,a,s) = 0.1$, and $\delta_1(t,a,s)=1$ for the first environment,
and $\delta_2(s,a,t) = 0.1, \delta_2(s,a,s) = 0.9$, and $\delta_2(t,a,s) = 1$ for the second environment.
Now, at state~$s$, repeating the action~$a$ a large number of times, and looking at the generated history,
one can guess with arbitrarily high confidence the current environment. However, no strategy can guess
the environment with certainty.
If, we rather set $\delta_1(s,a,t)=1$, and $\delta_2(s,a,s)= 1$, then an observed history uniquely determines
the current environment.

For the general {\sf SSP-ME} problem, there is an algorithm for an \emph{approximate} version of the above problem, namely the $\varepsilon$-gap problem. For any~$\varepsilon > 0$, a procedure for the \emph{$\varepsilon$-gap {\sf SSP-ME} problem} answers \textsf{Yes}
if the {\sf SSP-ME} problem has a solution; it answers \textsf{No} if the {\sf SSP-ME} problem has no solution
when each $\alpha_i$ is replaced with $\alpha_i-\varepsilon$; and answers either \textsf{Yes} or \textsf{No} otherwise.
Intuitively, such a procedure gives a correct answer on positive instances, and on instances that are
clearly too far (by $\varepsilon$) to be satisfiable. However, there is an uncertainty zone of size $\varepsilon$
on which the answer is not guaranteed to be correct. The algorithm is based on a reduction to the first order theory of the reals (see~\cite{RaskinS14}).

\begin{theorem}
  The {\sf SSP-ME} problem and the $\varepsilon$-gap {\sf SSP-ME} are \textsf{NP}-hard.
  For any~$\varepsilon>0$, there is a procedure for the $\varepsilon$-gap {\sf SSP-ME} problem.
\end{theorem}

\smallskip\noindent\textbf{Additional results.}
In~\cite{RS-fsttcs14}, we restricted our study to MDPs with \emph{two} environments, and considered reachability, safety, and parity objectives. We proved these problems to be decidable in polynomial time
for almost-sure and limit-sure conditions. The general quantitative case, i.e., arbitrary satisfaction probabilities is shown to be NP-hard already for two environments and MDPs with no cycles other than self-loops.
We gave a doubly exponential-space procedure to solve the $\varepsilon$-gap problem for reachability. 
We are currently studying the exact complexity of the case of arbitrary number of environments.

\section{Conclusion}
Through this paper, we gave an overview of classical approaches to the quantitative evaluation of strategies in MDPs, and presented three recent extensions that increase the modeling power of that framework. We chose to illustrate them through application to the stochastic shortest path problem. We hope this helps in understanding and comparing the different approaches. Let us sum up.

Given a weighted MDP modeling a stochastic shortest path problem, a first natural question is to find a strategy that minimizes the \textit{expected} sum of weights to target. This is the {\sf SSP-E} problem. Optimizing the average behavior of the controller is interesting if the process is to be executed a great number of times, but it gives no guarantee on individual runs, which may perform very badly. For a risk-averse controller, it may be interesting to look at the {\sf SSP-P} problem, which asks to maximize the \textit{probability} that runs exhibit an acceptable performance. When one really wants to ensure that no run will have an unacceptable performance, it is useful to resort to the {\sf SSP-G} problem, which asks to optimize the \textit{worst-case} performance of the controller.

In recent works, we introduced three related models that may be used to synthesize strategies with richer performance guarantees. First, if one reasons using the {\sf SSP-G} problem, he may obtain a strategy which is sub-optimal on average while using the {\sf SSP-E} problem gives no worst-case guarantee. With the framework of \textit{beyond worst-case synthesis}, developed in~\cite{BFRR-stacs14,DBLP:journals/corr/BruyereFRR14}, and presented here as the {\sf SSP-WE} problem, we can build strategies that provide \textit{both} worst-case guarantees and good expectation. Second, we are interested in describing rich constraints on the performance profile of strategies in multi-dimensional MDPs. To that end, we extended the {\sf SSP-P} problem to the {\sf SSP-PQ} problem, which handles multi-constraint \textit{percentile queries}~\cite{RRS14}. Those queries are particularly useful to characterize trade-offs between, for example, the length of a journey and its cost. Third and finally, we have discussed another extension of the {\sf SSP-P} problem that models some uncertainty about the stochastic model of the environment which is defined in the MDP through the transition function. With the {\sf SSP-ME} problem, we are able to analyze \textit{multi-environment} MDPs and synthesize strategies with guarantees against all considered environments~\cite{RS-fsttcs14,RaskinS14}.

\bibliographystyle{abbrv}
\bibliography{biblio-noh}
\end{document}